\documentclass[12pt,preprint]{aastex}
\usepackage{amsmath,bm}
\begin{document}
\title{Gamma-Ray Burst Optical Afterglows with Two-Component Jets: Polarization Evolution Revisited}
\author{Mi-Xiang Lan$^{1,2}$, Xue-Feng Wu$^{2,3}$, and Zi-Gao Dai$^{1,4}$}
\affil{$^{1}$School of Astronomy and Space Science, Nanjing University, Nanjing 210093, China; mxlan@pmo.ac.cn, dzg@nju.edu.cn \\
$^{2}$Purple Mountain Observatory, Chinese Academy of Sciences,
Nanjing 210008, China; xfwu@pmo.ac.cn \\
$^{3}$School of Astronomy and Space Science, University of Science and Technology of China,
Hefei, Anhui 230026, China\\
$^{4}$Key Laboratory of Modern Astronomy and Astrophysics (Nanjing
University), Ministry of Education, China \\}

\begin{abstract}
Gamma-ray bursts (GRBs) have been widely argued to originate from binary compact object mergers or core collapses of massive stars. Jets from these systems may have two components, an inner, narrow sub-jet and an outer, wider sub-jet. Such a jet subsequently interacts with its ambient gas, leading to a reverse shock (RS) and a forward shock (FS). The magnetic field in the narrow sub-jet is very likely to be mixed by an ordered component and a random component during the afterglow phase. In this paper, we calculate light curves and polarization evolution of optical afterglows with this mixed magnetic field in the RS region of the narrow sub-jet in a two-component jet model. The resultant light curve has two peaks: an early peak arises from the narrow sub-jet and a late-time rebrightening is due to the wider sub-jet. We find the polarization degree (PD) evolution under such a mixed magnetic field confined in the shock plane is very similar to that under the purely ordered magnetic field condition. The two-dimensional ``mixed'' magnetic fields confined in the shock plane are essentially the ordered magnetic fields only with different configurations. The position angle (PA) of the two-component jet can change gradually or abruptly by $90^\circ$. In particular, an abrupt $90^\circ$ change of the PA occurs when the PD changes from its decline phase to rise phase.

\end{abstract}


\section{Introduction}
Gamma-ray bursts (GRBs) are the most violent explosions at cosmological distances. Their central engines must be very powerful. Two kinds of central engines are involved. One is a black hole plus accretion disk system (Narayan, Paczy\'nski, \& Piran 1992; Woosley 1993; M\'esz\'aros \& Rees 1997; Paczy\'nski 1998). The other is a millisecond magnetar (Usov 1992; Duncan \& Thompson 1992; Klu\'zniak \& Ruderman 1998; Dai \& Lu 1998a,1998b; Spruit 1999; Ruderman, Tao, \& Klu\'zniak 2000; Wheeler et al. 2000). Jets launched from these central engines will pass through the stellar envelopes of massive stars or the ejecta of binary compact object mergers. Thus a cocoon is formed. The structure of the cocoon is a cone or a cylinder after the jet breaks out of the envelope or the ejecta (Morsony et al. 2007; Brombery et al. 2011; Mizuta \& Ioka 2013). As a phenomenological model, the two-component jet model (TCJM) with an ultra-relativistic narrow sub-jet and a mildly relativistic wider sub-jet can well describe the jet-cocoon structure.

This so-called TCJM has several other different meanings in the literature. First, Levinson \& Eichler (1993) proposed an outflow model, with a relativistic, baryon-poor inner sub-jet. Around the inner sub-jet is a subrelativistic, baryon-rich wind, which is driven by the disk. Second, a TCJM related to a black hole for both GRBs and active galactic nuclei was proposed by Xie et al. (2012). In this model, the inner sub-jet is driven by the Blandford-Znajek mechanism (BZ mechanism, Blandford \& Znajek 1977). The main composition of the inner sub-jet is Poynting flux, while the outer sub-jet, driven by the Blandford-Payne mechanism (BP mechanism, Blandford \& Payne 1982), is mainly composed by baryons. Their model gives rise to a spine/sheath jet structure. Third, Vlahakis et al. (2003) proposed a TCJM associated with a neutron star or a neutron-rich disk. In this model, a jet, composed of neutrons, protons and Poynting-flux, is initially accelerated. After it reaches a moderate Lorentz factor, neutrons are decoupled while protons and electrons keep on accelerating and collimating by electromagnetic forces. A highly collimated, relativistic proton-electron-dominated sub-jet and a wider, subrelativistic neutron-rich component, which finally decay to protons, are formed.

Huang et al. (2004) used the TCJM to explain the rapid rebrightening of the optical afterglow on the 14th day after the X-ray flash (XRF) 030723 and suggested that the TCJM provides a unified picture for GRBs and XRFs. The TCJM was also used to explain the light curves of GRB 030329, of which rebrightenings appear in both optical and radio band around 1.5 days after the burst (Berger et al. 2003). Wu et al. (2005) discussed the polarization evolution of GRB optical afterglows with the TCJM. In their paper, the dynamics of both sub-jets were assumed to follow the Blandford-McKee self-similar solution (Blandford $\&$ McKee 1976) and only the emission from forward shocks of both sub-jets with a random magnetic field configuration (MFC) was considered.

The MFC affects the polarization evolution significantly (Shaviv \& Dar 1995; Gruzinov \& Waxman 1999; Eichler \& Levinson 2003; Granot \& K\"{o}nigl 2003; Granot 2003; Lyutikov et al. 2003; Nakar et al. 2003; Dai 2004; Levinson \& Eichler 2004; Lazzati et al. 2004; Rossi et al. 2004; Wu et al. 2005; Lazzati 2006; Toma et al 2009; Beloborodov 2011; Inoue et al. 2011; Zhang \& Yan 2011; Lan, Wu \& Dai 2016a,b). The detailed numerical simulations show that the MFC is mixed after the prompt phase with both an ordered component remnant and a random component after magnetic dissipation (Deng et al. 2015). The polarization observations during the early optical afterglow phase suggest a polarization degree (PD) of $20\%-30\%$ (Mundell et al. 2013), which is not as high as $\sim70\%$ in the prompt phase (e.g., Yonetoku et al. 2012) and is also not zero. The moderate PD values during early afterglow phase are also confirmed by the numerical simulations (Deng et al. 2016,2017). This may indicate that the magnetic field in the jet is partly ordered during the afterglow phase. The polarization properties under such a mixed magnetic field should be considered.

There are 4 emission regions in the TCJM if the reverse shock (RS) regions of both sub-jets are included. Since the polarization evolution around the jet RS crossing time is very important (Lan, Wu \& Dai 2016a), a further discussion including both the RS contribution and a mixed MFC in the narrow sub-jet is needed. This paper is arranged as follows. In Section 2, the jet structure and MFCs in the corresponding jet are described. In Section 3, polarization properties with a mixed magnetic field are considered. In Section 4, numerical results are presented. In Section 5, we give our conclusions and discussion. In our calculation, a flat Universe with $\Omega_M=0.27$, $\Omega_\Lambda=0.73$ and $H_0=71{\rm \,km\,s^{-1}\,Mpc^{-1}}$ is adopted.

\section{Jet Structure and MFCs}
Two possible kinds of central engines for GRBs are black hole + accretion disk system and millisecond magnetar. Jets from these systems will interact with the stellar envelopes or the ejecta of binary compact object mergers, resulting in a jet-cocoon structure. TCJM as a phenomenological model of the jet-cocoon structure is considered here. In this paper, we assume that the distribution of both the Lorentz factor and the energy density are uniform in each sub-jet. The inner sub-jet is ultra-relativistic while the outer sub-jet is mildly relativistic. We also assume that there is no interaction between the two sub-jets. The lateral expansion is not considered for both sub-jets. Therefore, the dynamics of the two sub-jets are calculated separately using a forward-reverse-shock dynamics (Lan, Wu \& Dai 2016a). In our TCJM, there are four emission regions, i.e., the RS region of the narrow component, the forward shock (FS) region of the narrow component, the RS region of the wider component and the FS region of the wider component.

The narrow sub-jet with different origin mechanisms may have different ordered magnetic field components. The ordered magnetic field component in a jet powered by a black hole through the BZ mechanism is very likely to be toroidal but might be aligned in a jet driven by a magnetar (Spruit et al. 2001). For the aligned MFC in a GRB jet, according to the conservation of the magnetic flux, we have $\phi_{B_A}\sim R\theta_j\Delta B_A\sim {\rm const}$, where $R$ is the radius, $\theta_j$ is the half-opening angle of the jet, $\Delta$ is the width of the jet, $B_A$ is the strength of the aligned magnetic field at the shock plane. During the afterglow phase, the jet width $\Delta$ is roughly a constant when the reverse shock is crossing for the thick shell case while it increases as $R/\gamma^2$ for the thin shell case. Therefore, $B_A$ will decay no faster than $R^{-2}$ before the reverse shock crossing time. The radial component $B_r$ of the magnetic field decays as $R^{-2}$ at all times (Spruit et al. 2001; Zhang \& Kobayashi 2005). Therefore, the aligned component $B_A$ is dominating the radial component $B_r$ in the jet. In addition, during the prompt phase, because of some dissipation processes (e.g., collisions, shocks or magnetic reconnection), a tangled magnetic field would be generated. The resulting magnetic field is mixed with both an ordered remnant and a tangled component. For the wider jet (i.e., the cocoon), its mass is a mixture of the narrow sub-jet with the envelope or the ejecta. Therefore, the magnetic field in it is not likely to be large-scale ordered. We assume that it is random. Since the magnetic field in the interstellar medium (ISM) is not likely to be large-scale ordered, we assume that it is random in the FS region for both jets.

\section{Polarization Evolution with a \textbf{2-dimensional (2D)} mixed magnetic field}
In our previous studies (Lan, Wu \& Dai 2016a), we assume that the MFC in the RS region of the jet is entirely ordered. In reality, it is very likely to be mixed, with an ordered component carried out from the central engine and a random component generated by turbulence, magnetic reconnections or shocks. Here, we consider such a mixed magnetic field in the RS region of the narrow sub-jet and further assume that the mixed magnetic field is confined in the shock plane. The observed flux of the RS region in the narrow sub-jet can be expressed as
\begin{equation}
F_{\nu,rs,N}=\frac{1+z}{4\pi D_L^2}\frac{\sqrt{3}e^3}{m_ec^2}B'\int_0^{\theta_N+\theta_V} d\theta \mathcal{D}^3\sin\theta\int_{-\Delta\phi_N}^{\Delta\phi_N} d\phi\sin\theta'_B\int d\gamma_e N(\gamma_e)F(x),
\end{equation}
where $z$ is the redshift of the source, $D_L$ is the luminosity distance, $e$ and $m_e$ are the charge and mass of an electron, respectively. $c$ is the speed of the light. $B'$ is the strength of the total magnetic field $\vec{B}'$ in the jet comoving frame. $\theta$ is the angle between the line of sight (LOS) and velocity of the jet element (where we assume that the jet has no lateral expansion and the velocity of the jet element is radial). $\phi$ is the angle in the plane of sky between the projection of the jet axis and the projection of the velocity of the jet element. $\mathcal{D}=1/\gamma(1-\beta\cos\theta)$ is the Doppler factor. $\theta'_B$ is the pitch angle of the electrons, which is the angle between the velocity of the electrons and the direction of the total magnetic field $\vec{B}'$. $N(\gamma_e)$ is the energy spectrum of the electrons. $F(x)$ is dimensionless synchrotron spectrum. $x=\nu'/\nu'_c$ with $\nu'=\nu_{obs}(1+z)/\mathcal{D}$ and $\nu'_c=eB'\sin\theta'_B\gamma_e^2/2\pi m_ec$. And the corresponding Stokes parameters $Q_{\nu,3}$ and $U_{\nu,3}$ can be expressed as
\begin{equation}
Q_{\nu,rs,N}=\frac{1+z}{4\pi D_L^2}\frac{\sqrt{3}e^3}{m_ec^2}B'\int_0^{\theta_N+\theta_V} d\theta \mathcal{D}^3\sin\theta\int_{-\Delta\phi_N}^{\Delta\phi_N} d\phi \sin\theta'_B\cos(2\chi_f)\int d\gamma_e  N(\gamma_e)G(x),
\end{equation}
\begin{equation}
U_{\nu,rs,N}=\frac{1+z}{4\pi D_L^2}\frac{\sqrt{3}e^3}{m_ec^2}B'\int_0^{\theta_N+\theta_V} d\theta \mathcal{D}^3\sin\theta\int_{-\Delta\phi_N}^{\Delta\phi_N} d\phi \sin\theta'_B\sin(2\chi_f)\int d\gamma_e  N(\gamma_e)G(x),
\end{equation}
where $G(x)=xK_{2/3}(x)$ with $K_{2/3}(x)$ being the modified Bessel function of 2/3 order. $\chi_f$ is the local position angle (PA) of the emission with the total magnetic field $\vec{B}'$.

In the following, we derive an expression of the local polarization PA $\chi_f$. We establish two right-handed orthogonal coordinate systems in the local point-like region (where the comoving observational direction is fixed), $\hat{x}\hat{y}\hat{\beta}$ and $\hat{1}\hat{2}\hat{k}'$, which are shown in Fig. 1 (also see Sari 1999; Toma et al. 2009 ). $\hat{\beta}$ is the direction of the bulk velocity of the jet element and $\hat{y}$ is parallel to the direction of $\hat{\beta}\times\hat{k}$, where $\hat{k}$ represents the direction of the LOS. Let $\hat{k}'$ be the comoving LOS and $\hat{1}=\hat{y}$. $\eta'$ is the azimuthal angle of the total magnetic field $\vec{B}'$ in the coordinate system $\hat{x}\hat{y}\hat{\beta}$. $\theta_B'$ and $\phi_B'$ are the polar and azimuthal angles of the total magnetic field $\vec{B}'$ in the coordinate system $\hat{1}\hat{2}\hat{k}'$. Comparing the components of $\hat{B}'$ in two coordinate systems, we get the following relations (Toma et al. 2009)
\begin{align}
\cos\theta_B'&=\cos\eta'\sin\theta',  \nonumber \\
\cos\phi_B'&=\sin\eta'/\sin\theta_B', \nonumber \\
\sin\phi_B'&=-\cos\eta'\cos\theta'/\sin\theta_B'.
\end{align}
Notice that the expression of $\sin\phi'_B$ in Toma et al. (2009) lacks a minus sign. Then in the coordinate system $\hat{1}\hat{2}\hat{k}'$, the electric vector satisfies $\hat{e}'\parallel \hat{B}'\times\hat{k}'$. We have $\hat{e}'=\cos\chi\hat{1}+\sin\chi\hat{2}$ with $\chi=\phi_B'-\pi/2$. We now express the coordinate axis $\hat{1}$ and $\hat{2}$ in the global coordinate system $\hat{X}\hat{Y}\hat{k}$ with $\hat{k}$ the LOS and $\hat{X}$ being the projection of the jet axis on the plane of the sky.
\begin{align}
\hat{1}&=\hat{\beta}\times\hat{k}/|\hat{\beta}\times\hat{k}|=(\sin\phi, -\cos\phi, 0),  \nonumber \\
\hat{2}&=\hat{k}'\times\hat{1}=\mathcal{D}\left(\cos\phi(1+A\cos\theta), \sin\phi(1+A\cos\theta), -A\sin\theta\right).
\end{align}
where we denote $A\equiv(\gamma-1)\cos\theta-\gamma\beta$. Then we transform the electric vector to the observer frame,
\begin{equation}
\vec{e}=\mathcal{D}\hat{e}'-(\hat{\beta}\cdot\hat{e}')[(\gamma-1)\hat{\beta}+\gamma\beta\hat{k}'].
\end{equation}
It is easy to confirm $e_k=0$. We then have
\begin{equation}
\chi_f=\arctan\left(\frac{e_Y}{e_X}\right)=\arctan\left(-\cot(\chi+\phi)\right)=\phi'_B+\phi.
\end{equation}

As mentioned above, the magnetic field in the RS region of the narrow sub-jet is a mixed field, i.e., $\vec{B}'=\vec{B}'_{ord}+\vec{B}'_{rnd,rs,N}$ with $\vec{B}'_{ord}$ and $\vec{B}'_{rnd,rs,N}$ being the ordered component and the random component, respectively. Here, the magnitude of the random magnetic field is given by $B'_{rnd,rs,N}=\sqrt{8\pi\epsilon_{B,rs,N}e'_{3,N}}$. $e'_{3,N}$ is the internal energy density of the RS region in the narrow sub-jet. A fraction $\epsilon_{B,rs,N}$ of the internal energy in the RS region goes into the random magnetic field. It is assumed to be \textit{$B'_{ord}\equiv\xi_B B'_{rnd,rs,N}$} for the ordered component. In a point-like region, the direction of the ordered magnetic field is fixed. For a toroidal MFC, we have the expression of its direction in the coordinate system $\hat{x}\hat{y}\hat{\beta}$: $\hat{B}'_T=(-J_{T,y}/A_T, J_{T,x}/A_T, 0)$. We denote $A_T=\sqrt{J_{T,x}^2+J_{T,y}^2}$ with $J_{T,x}=-\sin\theta_V\cos\theta\cos\phi+\cos\theta_V\sin\theta$ and $J_{T,y}=\sin\theta_V\sin\phi$. For an aligned MFC, its direction in the coordinate system $\hat{x}\hat{y}\hat{\beta}$ can be expressed as: $\hat{B}'_A=(-J_{A,y}/A_A, J_{A,x}/A_A, 0)$. We denote $A_A=\sqrt{J_{A,x}^2+J_{A,y}^2}$ with $J_{A,x}=\sin\delta_a\cos\theta_V\cos\theta\cos\phi+\cos\delta_a\cos\theta\sin\phi+\sin\theta_V\sin\delta_a\sin\theta$ and $J_{A,y}=-\sin\delta_a\cos\theta_V\sin\phi+\cos\delta_a\cos\phi$. Here, we assume that the aligned magnetic fields are latitude circles with axis $\hat{J_A}$, where $\hat{J_A}\perp\hat{J}$. $\delta_a$ is the angle between $J_{A}$ and the vector $\hat{J}\times\hat{k}$. Eq. (7) is consistent with Eq. (23) of our previous paper (Lan, Wu \& Dai 2016a). If we set the random component of the magnetic field is zero, then $(\cos\eta',\sin\eta',0)=\hat{B}'_T$, and we get $\tan\phi'_B=\sin\phi'_B/\cos\phi'_B=-\cos\theta'\cos\eta'/\sin\eta'=\cos\theta'J_{T,y}/J_{T,x}$. Finally, $\chi_f=\phi+\phi'_B=\phi+\arctan(\cos\theta'J_{T,y}/J_{T,x})$, which is same as Eq. (23) of Lan, Wu \& Dai (2016a) after $J_{T,x}$ and $J_{T,y}$ are taken into account.

We assume that the ordered and random components of the magnetic field are both confined in the shock plane. We denote $\eta'_{rnd}$ is the azimuthal angle of the random magnetic field in the coordinate system $\hat{x}\hat{y}\hat{\beta}$. Here, we find that the random magnetic field is perpendicular to the direction of the ordered magnetic field because of $\vec{B}'=\vec{B}'_{ord}+\vec{B}'_{rnd,rs,N}$ and $B^{'2}=B^{'2}_{ord}+B^{'2}_{rnd,rs,N}$, \textbf{leading to \boldmath$\vec{B}'_{ord}\cdot\vec{B}'_{rnd,rs,N}=0$\unboldmath}. Therefore, we have in the shock plane $(\cos\eta'_{rnd},\sin\eta'_{rnd})=(B'_y,-B'_x)$, where $B'_x$ and $B'_y$ denote the x- and y-components of the unit vector of the ordered magnetic field component $\hat{B}'_{ord}$.

Finally, the PD ($\Pi_{TCJ}$) and PA ($\chi_{TCJ}$) of the emission from the two-component jet (TCJ) can be expressed as
\begin{equation}
\Pi_{TCJ}=\frac{\sqrt{Q_{\nu,TCJ}^2+U_{\nu,TCJ}^2}}{F_{\nu,TCJ}}
\end{equation}
\begin{equation}
\chi_{TCJ}=\frac{1}{2}\arctan\frac{U_{\nu,TCJ}}{Q_{\nu,TCJ}}
\end{equation}
where $F_{\nu,TCJ}=\sum\limits_i\sum\limits_j F_{\nu,i,j}$, $Q_{\nu,TCJ}=\sum\limits_i\sum\limits_j Q_{\nu,i,j}$ and $U_{\nu,TCJ}=U_{\nu,rs,N}$ are the total Stokes parameters from the TCJ with $i=rs$ for the RS region, $i=fs$ for the FS region, $j=N$ for the narrow sub-jet and $j=W$ for the wider sub-jet. The expression of the Stokes parameters with random magnetic field in the emission region can be found in our previous paper (Lan, Wu \& Dai 2016a). When calculating the polarization evolution with the random magnetic field, the spectral index $m$ ($F_\nu\propto\nu^m$) is needed. Here, we use the formula $m=\ln(F_{\nu_1}/F_{\nu})/\ln(\nu_1/\nu)$ with $\nu_1=\nu+\Delta\nu$, where $\Delta\nu$ is a small change of the observational frequency $\nu$. The integral over the azimuthal angle $\phi$ (the angle in the plane of the sky between the projection of the jet axis and the projection of the velocity of the jet element) of the wider sub-jet has two ranges: ($-\Delta\phi_W$, $-\Delta\phi_N$) and ($\Delta\phi_N$, $\Delta\phi_W$), with $\Delta\phi_i$ given in our previous papers (Lan, Wu \& Dai 2016a; Wu et al. 2005). The integral range of $\theta$ for the wider sub-jet is from 0 to $\theta_W+\theta_V$.

The evolution of the PA is determined by the formula $\chi=1/2\arctan(U_\nu/Q_\nu)$, if both $Q_\nu$ and $U_\nu$ are nonzero. But the actual value of PA cannot be obtained by the only use of this formula. Additionally, we need the sign of Stokes parameters $Q_\nu$ and $U_\nu$ to obtain the real PA value. Namely, we get a PA value from the formula $\chi=1/2\arctan(U_\nu/Q_\nu)$. If $Q_\nu >0$, then the actual PA value $\chi_r$ equals to $\chi$. If $Q_\nu <0$ and $U_\nu >0$, then the actual PA value $\chi_r$ is equal to $\chi+\pi/2$. If $Q_\nu <0$ and $U_\nu <0$, however, then the actual PA value $\chi_r$ is equal to $\chi-\pi/2$. We consider these for PA evolution in this paper.

\section{Numerical Results}
\subsection{Dynamics}
For the narrow sub-jet, two kinds of dynamics are considered. One corresponds to the thin shell case, the other is for the thick shell case (Sari \& Piran 1995). The dynamical parameters we take for the thin shell case are as follows: $E_{iso,N}=10^{51}\,{\rm ergs}$, $\eta_N=200$, $\Delta_{0,N}=10^{10}\,{\rm cm}$ and $\theta_N=0.03$ rad. The parameters we take for the thick shell case are: $E_{iso,N}=10^{51}\,{\rm ergs}$, $\eta_N=300$, $\Delta_{0,N}=10^{12}\,{\rm cm}$ and $\theta_N=0.03$ rad. For the wider sub-jet, its initial Lorentz factor is lower by about one order of magnitude than that of the narrow sub-jet, so we only consider the newtonian RS, i.e. the thin shell case. The dynamic parameters we take for the wider sub-jet are: $E_{iso,W}=10^{52}\,{\rm ergs}$, $\eta_W=15$, $\Delta_{0,W}=10^{12}\,{\rm cm}$, $\theta_W=0.3$ rad. $E_{iso,j}$ is the isotropic equivalent energy, $\eta_j$ is the initial Lorentz factor, $\Delta_{0,j}$ is the initial width of the jet, $\theta_j$ is the half-opening angle of the jet. For the wider sub-jet, it is a hollow cone with inner edge $\theta_N$ and outer edge $\theta_W$. The source is assumed to be located at redshift of $z=0.1$ with an ISM density $n_1=1\,{\rm cm}^{-3}$. The RS crossing time $t_{c,N}$ of the narrow sub-jet is 12.8 s for the thin shell case and 24.4 s for the thick shell. The RS crossing time $t_{c,W}$ for the wider sub-jet is $2.75\times10^4$ s. The dynamics of the both jets considered in this paper are shown in Fig. 2.

\subsection{The Flux Ratio of RS to FS Regions}
Because the PD of the synchrotron emission is usually high in the ordered magnetic field and the large-scale ordered component of the magnetic field may exist in the RS region of the jet, it is necessary to consider under what conditions the RS emission will dominate the total flux. Whether the RS emission dominates the total observed flux or not depends on several parameters, especially the $R_B$ and $\xi_B$ factors. The $R_B$ factor is defined as \textit{$R_B\equiv\epsilon_{B,rs,N}/\epsilon_{B,fs,N}$} (Zhang, Kobayashi \& M\'esz\'aros 2003). $\epsilon_{B,fs,N}$ is the energy participation factor of the random magnetic field in the FS region of the narrow sub-jet. The value of the energy participation factor of the magnetic field in the FS region $\epsilon_{B,fs,j}$ is very low (Kumar et al. 2010; Wang et al. 2015). Here, we take $\epsilon_{B,fs,N}=10^{-5}$. We consider the dependence of the ratio \textit{$\zeta\equiv f_{\nu,rs,N}(t_{c,N})/f_{\nu,fs,N}(t_{c,N})$} of the RS flux to the FS flux at the RS crossing time of the narrow sub-jet $t_{c,N}$ on the two parameters mentioned above. The dynamics used here for the thin and thick shell cases of the narrow sub-jet are shown in Fig. 2. The other fixed emission parameters for the narrow sub-jet we take are as follows: $\epsilon_{e,rs,N}=\epsilon_{e,fs,N}=0.1$, $p_{rs,N}=p_{fs,N}=2.5$. $\epsilon_{e,i,N}$ is the energy participation factor of the electrons in the region $i$ of the narrow sub-jet. $p_{i,N}$ is the spectral index of the injected electrons in region $i$. The observational angle is 0. The results for the thin and thick shell cases are shown in Fig. 3. For both the thin and thick shell cases, under the fixed parameters we take, the value of $\zeta$ increases with $R_B$ ($\xi_B$ fixed), while it keeps as a constant until $\xi_B\sim1.0$ and then increases with $\xi_B$ ($R_B$ fixed).

\subsection{Polarization Evolution}
We first consider the emission from the RS region of the narrow sub-jet. We take a set of parameters as the ``fiducial set'' (Case (1)) with  $\xi_B=10$, thin shell, the toroidal ordered magnetic field component and $\theta_V=0.6\theta_N$. Once change a parameter ($\xi_B=0.1$ for Case (2), $\xi_B=1$ for Case (3), aligned ordered magnetic field component for Case (4), thick shell for Case (5) and $\theta_V=3\theta_N$ for Case (6)) to discuss its effects on the polarization properties. The fixed emission parameters for all 6 cases we take are $\epsilon_{B,rs,N}=10^{-5}$, $\epsilon_{e,rs,N}=0.1$, and $p_{rs,N}=2.5$. The orientation of the aligned magnetic field (if there is) is taken to be $\delta_a=\pi/6$. The dynamics used for the narrow sub-jet of the thin and thick shell cases is shown in Fig. 2. The results of light curves and polarization evolution are shown in Fig. 4. Because there are no fresh electrons in the RS region after the RS crossing time, the flux will drop exponentially once $\nu>\nu_{cut}$ (Kobayashi 2000; Zou et al. 2005). We then set it to be zero. The MFCs in the visible region (i.e., the $1/\gamma$ cone) are approximate to be aligned before and slightly after the RS crossing time for all cases except for Case (6). This will result in constant PD values for these cases at the beginning. For Cases (1), (2), (3) and (5) (the toroidal ordered magnetic field component cases), the PD values begin to decrease slightly after the RS crossing time. The bulk Lorentz factor of the narrow sub-jet decreases sharply after the RS crossing time. More and more complete circles of the magnetic field enter the increasing visible region leading to the decrease of the asymmetry and then of the PD values. For Cases (1), (2) and (3) (corresponding to $\xi_B=10,\ 0.1,\ 1$), the PD values are almost the same as the time evolution. The mixed magnetic fields considered here are assumed to be confined in the shock plane which causes the random magnetic field component having a fixed direction perpendicular to the ordered component in the point-like region. Different values of $\xi_B$ will correspond to different MFCs in the shock plane and these different MFCs are all essentially large-scale ordered. Furthermore, under parameters we take, both the polarized flux and the total flux increase with the value of $\xi_B$. Their ratios (i.e., the PD values) are almost the same for different $\xi_B$ values. For Case (4) (i.e., the aligned ordered magnetic field component case), the PD begins to increase slightly after the RS crossing time. With the increase of the $1/\gamma$ cone after the RS crossing time, the visible region will not be covered by the jet region and a new asymmetry appears, which leads to an increase of the the PD value. For Case (6), the PD begins to rise after 100 s and reaches its maximum value when $1/\gamma\sim\theta_V-\theta_N$ (Waxman 2003). The last PD value of Case (1) rises to 0.34 because the flux contribution from the large $\theta$ values vanishes\footnote{Because of the Doppler boosting, the $\nu'(\theta)$ ($\nu'(\theta)=\nu_{obs}(1+z)\gamma(1-\beta\cos\theta)$) increases with $\theta$, at some critical $\theta_0$, $\nu'(\theta_0)=\nu'_{cut}$, then when $\theta_0<\theta\leq\theta_j+\theta_V$, we will have $\nu'(\theta)>\nu'_{cut}$ and the flux from these regions vanishes.}, leading to less complete magnetic field circles in the emitting region (i.e., increasing the asymmetry in the emission region). The PA of Case (2) changes by roughly $90^\circ$ around 3000 s, while it changes gradually by approximately $45^\circ$ around 1000 s in Case (3). The PA values in the other cases are roughly constant.

We next calculate the light curves and polarization evolution for the narrow sub-jet including the emissions from both the RS and FS regions, which are shown in Fig. 5. The Stokes parameters of the emission from the RS region used in Fig. 5 are the same as those in Fig. 4. For the FS region, the emission parameters that we take are as follows: $\epsilon_{B,fs,N}=10^{-5}$, $\epsilon_{e,fs,N}=0.1$ and $p_{fs,N}=2.5$. For the thin shell cases, if the flux from the RS region dominates the total flux (Cases (1) and (4)), there are bumps in PD evolution around the RS crossing time and the peak value reaches about $50\%$, which is consistent with our previous study (Lan, Wu \& Dai 2016a). For Cases (2) and (3), the flux ratio of the FS (with lower PD values) to RS emissions is higher compared to that is Case (1). Therefore, the PD value is smaller during the RS crossing. For the thick shell case (i.e., Case (5)), the PD keeps roughly as a constant before the RS crossing time, which is also consistent with the result in our previous paper (Lan, Wu \& Dai 2016a). For Cases (1), (2), (3) and (5) (i.e., the toroidal ordered magnetic field component cases), the PD value decreases more quickly after the RS crossing time in Fig. 5 than that in Fig. 4 because the emission with the low PD value from the FS region becomes more and more important after $t_{c,N}$. For Case (4) (i.e., the aligned ordered magnetic field component case), the decrease of PD value after $t_{c,N}$ is also because the increasing flux from the FS region. For Case (6) (i.e., the off-axis observation), there is a bump in the PD evolution at late time, which is also consistent with our former study (Lan, Wu \& Dai 2016a). The PA for Case (4) (i.e., the aligned ordered magnetic field component) can change gradually. The change of PA can be roughly $45^\circ$ for Case (3). The abrupt changes of PA by $180^\circ$ in Fig. 5 may be not real and due to the mathematical definition. Most of the abrupt $90^\circ$ changes of the PA happen during the PD changing from decrease to rise for the narrow sub-jet. A spike of the PA in Cases (1) and (5) happens just before the flux from the RS region of the narrow sub-jet becomes zero. At this time, the Stokes parameter $U_{\nu,rs,N}$ changes its sign and the Stokes parameter $Q_{\nu}$ from the narrow sub-jet is less than zero. In addition, $\mid U_{\nu,rs,N}\mid\ll \mid Q_{\nu}\mid$, according to our analysis in Section 3, so there is a $\sim180^\circ$ spike in the PA evolution curve.

Finally, the total light curves and polarization evolution of the emission from the TCJ are presented in Fig. 6 including the contributions from 4 emission regions (i.e., the RS region of the narrow sub-jet, the FS region of the narrow sub-jet, the RS region of the wider sub-jet and the FS region of the wider sub-jet). The dynamics used for the wider sub-jet is shown in Fig. 2. The Stokes parameters of the emission from the narrow sub-jet in Fig. 6 are the same as that used in Fig. 5. The emission parameters we take for the wider sub-jet are $\epsilon_{e,rs,W}=\epsilon_{e,fs,W}=0.1$, $\epsilon_{B,rs,W}=\epsilon_{B,fs,W}=10^{-5}$ and $p_{rs,W}=p_{fs,W}=2.5$. Because both the flux and the polarized flux contributions from the wider sub-jet can be neglected during the early stage, the light curves and polarization properties are almost the same as that shown in Fig. 5 of which only the emission from the narrow sub-jet are considered. There are bumps in the light curves around the RS crossing time of the wider sub-jet $t_{c,W}\sim2.75\times10^4$ s. The changes of PA can be either gradually or abruptly for the TCJ. A spike of the PA $\sim10^4$ s after the burst in Case (5) of Fig. 6 disappears because the Stokes parameter $Q_\nu$ is dominated by the positive value from the wider sub-jet and $\mid U_{\nu,rs,N}\mid\ll \mid Q_{\nu}\mid$. Although the Stokes parameter $U_{\nu,rs,N}$ changes its sign, the PA hardly evolves with time. Most of the abrupt changes of the PA value seem to happen when the PD value changes from its decline phase to rise phase for the TCJ.

\section{Conclusions and Discussion}

Because the MFCs affect polarization evolution significantly and the mixed MFC is very likely to exist in the jet during the early afterglow phase, we have considered the polarization evolution with such a mixed MFC and then applied our model to the TCJM which is a phenomenal model of the jet-cocoon structure.

We assumed that there is no interaction between the two sub-jets. Thus, the hydrodynamic evolution can be considered separately. The initial Lorentz factor of the narrow sub-jet is assumed one order of magnitude larger than that of the wider sub-jet. The radii of the sub-jets, $R_i\sim c\gamma_i^2t_{obs}$, are different at the same observed time $t_{obs}$. The contributions to the Stokes parameters from the RS regions of both jet components are also considered.

In this paper, we considered a 2-dimensional mixed magnetic field which is assumed to be confined in the shock plane. We found that the random component has a fixed direction perpendicular to the ordered component. Our results of light curves and PD evolution in such a 2-dimensional ``mixed'' magnetic field are very similar to those of the purely ordered magnetic fields discussed in our previous paper (Lan, Wu \& Dai 2016a). In Fig. 5 (i.e., the light curves and polarization properties of the narrow sub-jet are shown), if the RS emission dominates over the FS emission, for the thin shell cases (e.g., Cases (1) and (4)) there are bumps in PD evolution at the RS crossing time while it keeps roughly as a constant before the RS crossing time for the thick shell case (Case (5)). The peak value of the PD bump in Cases (1) and (4) and the constant PD value before $t_{c,N}$ of Case (5) is roughly 50\%. The above results are consistent with our previous study (Lan, Wu \& Dai 2016a).

There are two peaks in the light curves of the TCJ with the early peak from the narrow sub-jet and the late-time rebrightening due to the wider sub-jet. The polarization properties at the early stage (around the RS crossing time of the narrow sub-jet) are mainly determined by the narrow sub-jet. The flux from the wider sub-jet becomes important around $t_{c,W}$. Since the MFC in the wider sub-jet is random, its contribution to the polarized flux is relatively low. Therefore, there is no bump in the PD evolution at $t_{c,W}$. The change of PA from the TCJ can be abruptly or gradually. The abrupt $90^\circ$ change of the PA happens during the PD changing from its decline phase to rise phase. The abrupt changes of the PA by $180^\circ$ may be due to the mathematical definition and does not necessarily mean that the direction of the magnetic field in the emission region changes by $180^\circ$.

\textbf{Finally, what we would point out is that the 2D ``mixed'' magnetic field adopted in this paper is essentially large-scale-ordered. The PD properties with this field are very similar to those with a purely-ordered magnetic field. In reality, the random magnetic field component could be generated by shocks, turbulence or magnetic reconnection. Therefore, a 3-dimensional (3D) mixed magnetic field (including both ordered and random components) should be more realistic. The recent numerical simulations show that, for a 3D magnetic field, the PD decreases with increasing the randomness of the magnetic field (Deng et al. 2016). Within the frame of this work, we will study the polarization properties with such a 3D mixed magnetic field elsewhere.}

\acknowledgements
We thank an anonymous referee for constructive suggestions that have allowed us to improve the manuscript significantly. This work is supported by the National Basic Research Program (``973'' Program) of China (grant No. 2014CB845800), the National Key Research and Development Program of China (grant No. 2017YFA0402600), and the National Natural Science Foundation of China (grant Nos. 11573014, 11673068 and 11725314). X.F.W is also partially supported by the Youth Innovation Promotion Association (2011231), the Key Research Program of Frontier Sciences (QYZDB-SSW-SYS005) and the Strategic Priority Research Program ``Multi-waveband gravitational wave Universe'' (grant No. XDB23040000) of the Chinese Academy of Sciences. M.X.L is also supported by the Natural Science Foundation of Jiangsu Province (Grant No. BK20171109).

\begin{figure}
\begin{center}
\includegraphics[width=1.0\textwidth,angle=0]{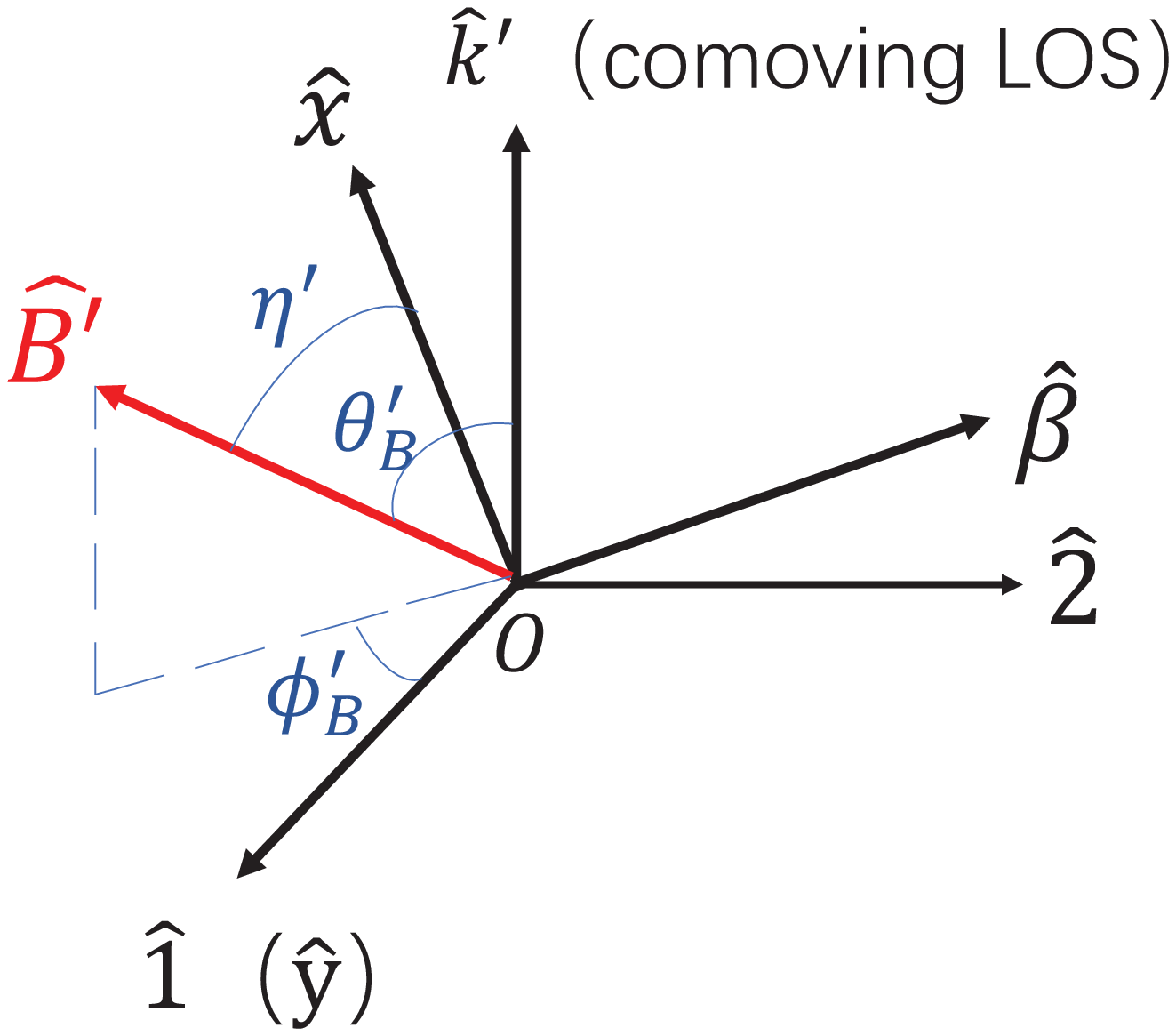}
\caption{A sketch of the coordinate systems used in our calculation. $\hat{\beta}$ is the local velocity direction of the jet element. $\hat{k}'$ is the comoving LOS, i.e., the comoving wave-vector. $\hat{B}'$ is the direction of the total magnetic field, which is assumed to be confined in the shock plane (i.e., in the $\hat{x}\hat{y}$ plane). $\theta'_B$ and $\phi'_B$ are the polar and azimuthal angle of $\hat{B}'$ in coordinate system $\hat{1}\hat{2}\hat{k}'$. $\eta'$ is its azimuthal angle in coordinate system $\hat{x}\hat{y}\hat{\beta}$.} \label{fig1}
\end{center}
\end{figure}

\begin{figure}
\begin{center}
\includegraphics[width=1.0\textwidth,angle=0]{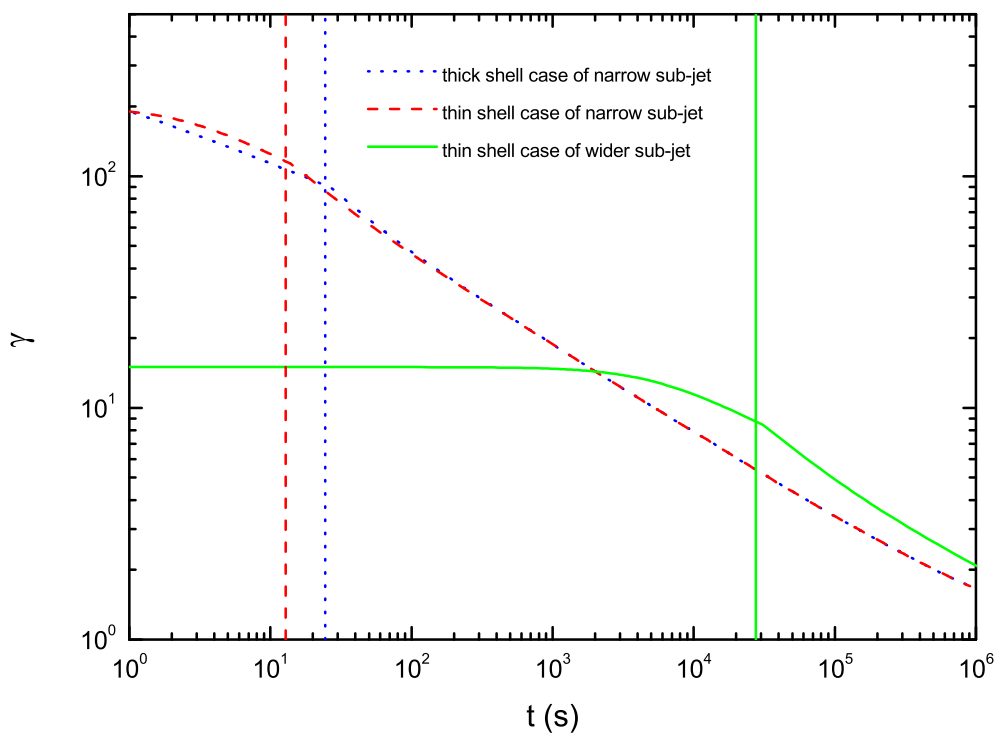}
\caption{Evolutions of the bulk Lorentz factors of the sub-jets. The blue dotted and red dashed lines correspond to the thick shell and thin shell cases of the narrow sub-jet, respectively. The green solid line is for the thin shell case of the wider jet. The vertical lines correspond to the RS crossing time, with left, medium and right ones for the thin shell case of the narrow sub-jet, thick shell case of the narrow sub-jet and the thin shell case for the wider sub-jet.} \label{fig2}
\end{center}
\end{figure}

\begin{figure}
\begin{center}
\includegraphics[width=1.0\textwidth,angle=0]{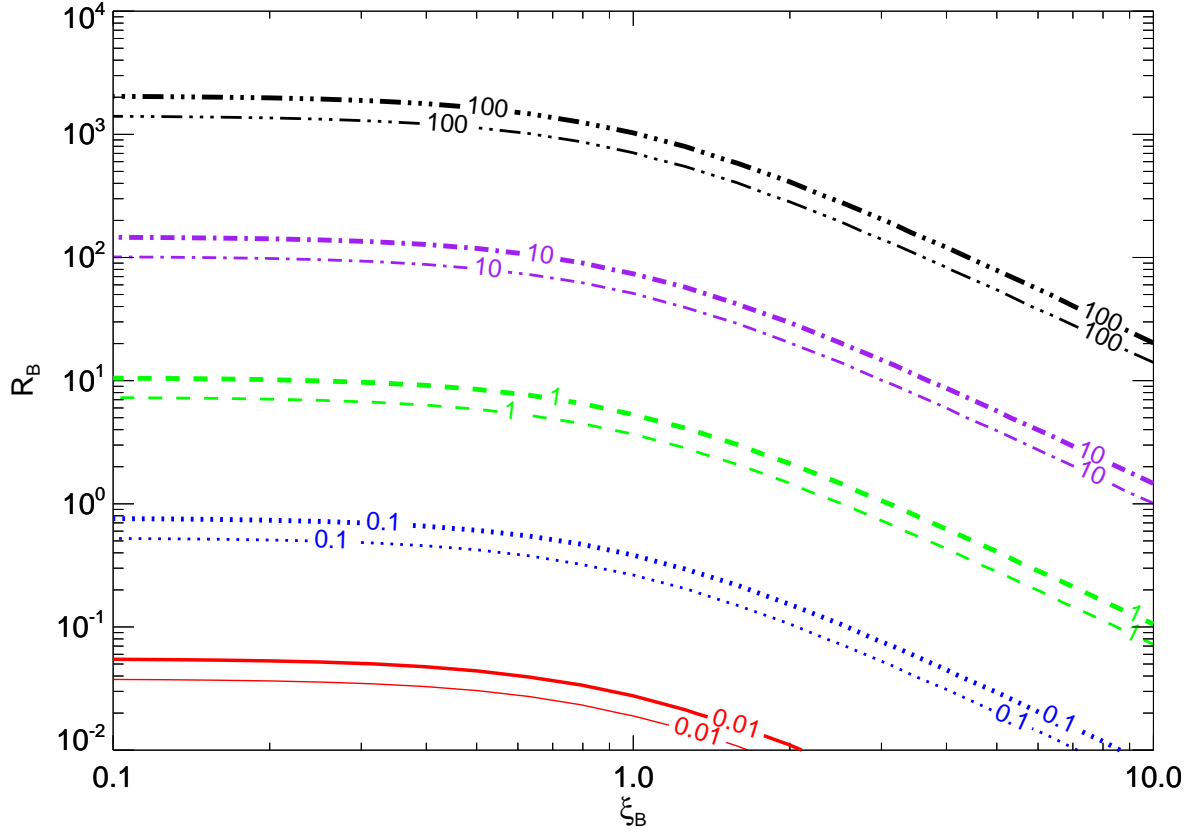}
\caption{Dependence of the flux ratio $\zeta$ on the $R_B$ and $\xi_B$ of the narrow sub-jet. The thick lines correspond to the thick shell case while the thin lines are related to the thin shell case. The solid, dotted, dashed, dash-dotted and dash-dot-dot-dot lines correspond to $\zeta=0.01, 0.1, 1, 10, 100$, respectively.} \label{fig3}
\end{center}
\end{figure}

\begin{figure}
\begin{center}
\includegraphics[width=1.0\textwidth,angle=0]{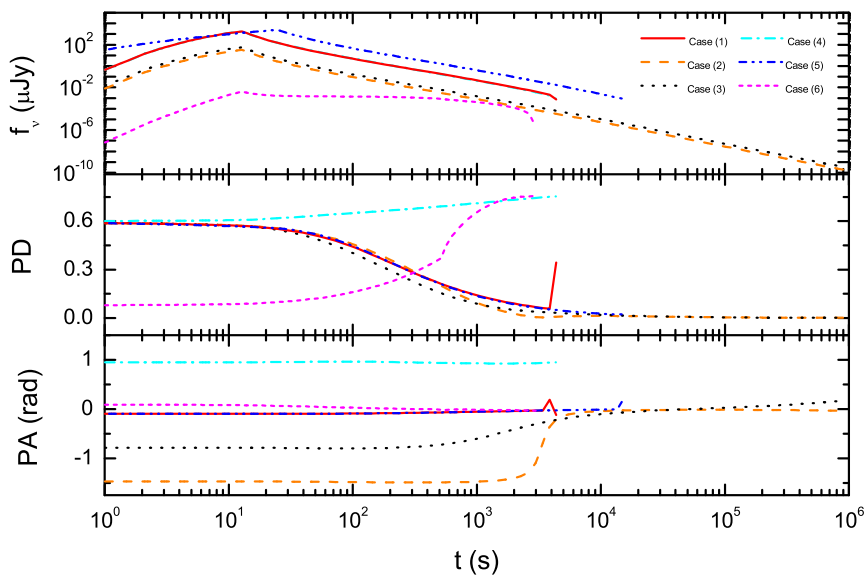}
\caption{The light curves and polarization evolutions of the emission from the RS region of the narrow sub-jet. The upper panel shows the light curves. The medium and the lower panels show the PD and PA evolutions, respectively. The red solid line corresponds to the basic parameter set, i.e., Case (1). The dashed, dotted, dash-dot, dash-dot-dot and short-dash lines correspond to Cases (2), (3), (4), (5) and (6).} \label{fig4}
\end{center}
\end{figure}

\begin{figure}
\begin{center}
\includegraphics[width=1.0\textwidth,angle=0]{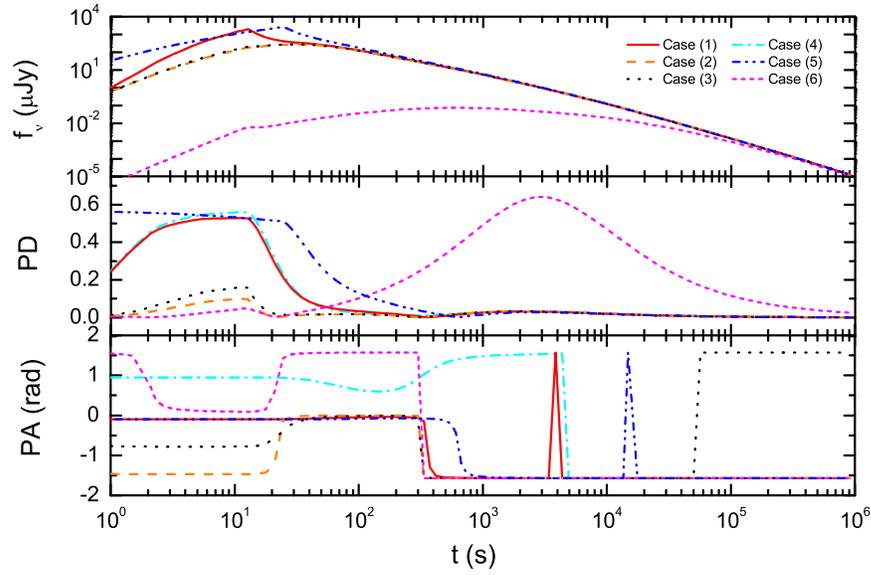}
\caption{Same as Fig. 4, but for the emission from the narrow sub-jet, including the contributions from both the RS and FS regions. } \label{fig5}
\end{center}
\end{figure}

\begin{figure}
\begin{center}
\includegraphics[width=1.0\textwidth,angle=0]{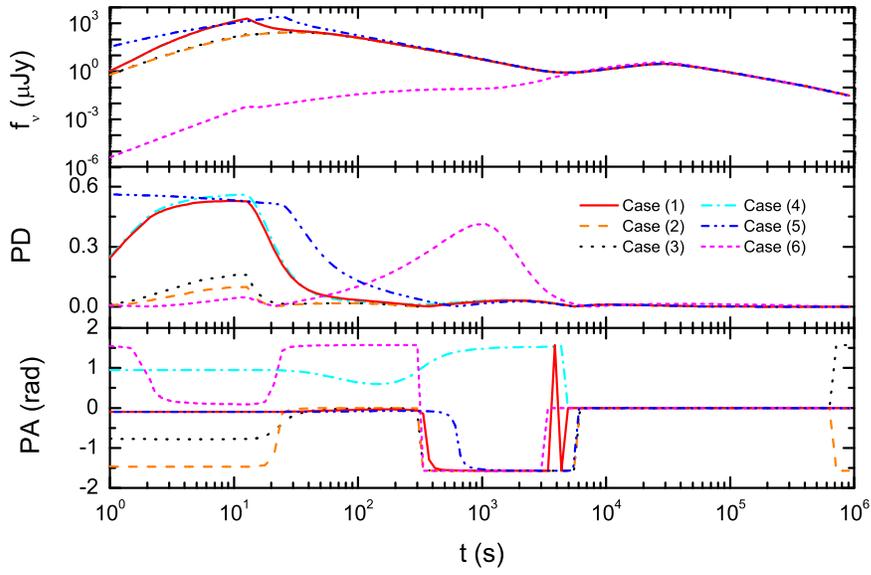}
\caption{Same as Fig. 4, but for the emission from the TCJ, including the contributions from both the narrow and wider sub-jets. } \label{fig6}
\end{center}
\end{figure}

\end{document}